\documentclass[pra,superscriptaddress, nofootinbib,showpacs]{revtex4}
\usepackage{graphicx}       
\usepackage{bm}         
\usepackage{amsfonts}
\usepackage{latexsym}
\usepackage{amssymb}

\begin{document}

\title[Security of QKD using Franson interferometers]{Security of high-dimensional quantum key distribution protocols using Franson interferometers}
\author{Thomas Brougham$^1$, Stephen M. Barnett$^1$, Kevin T. McCusker$^{2,3}$, Paul G. Kwiat$^2$ and Daniel J. Gauthier$^4$}
\address{$^1$ Department of Physics, University of Strathclyde, Glasgow, G4 0NG, U.K.\\
$^2$ Department of Physics, University of Illinois at Urbana-Champaign, Urbana, Illinois 61801, USA \\
$^3$ Department of Electrical Engineering and Computer Science, Northwestern University,\\2145 Sheridan Road, Evanston, IL, 60208-3118, USA\\
$^4$ Department of Physics, Duke University, Box 90305, Durham, North Carolina 27708, USA}

\begin{abstract}
Franson interferometers are increasingly being proposed as a means of securing high-dimensional energy-time entanglement-based quantum key distribution (QKD) systems.  Heuristic arguments have been proposed that purport to demonstrate the security of these schemes.  We show, however, that such systems are vulnerable to attacks that localize the photons to several temporally separate locations.  This demonstrates that a single pair of Franson interferometers is not a practical approach to securing high-dimensional energy-time entanglement based QKD.  This observations leads us to investigate the security of modified Franson-based-protocols, where Alice and Bob have two or more Franson interferometers.  We show that such setups can improve the sensitivity against attacks that localize the photons to multiple temporal locations.  While our results do not constituting a full security proof, they do show that a single pair of Franson interferometers is not secure and that multiple such interferometers could be a promising candidate for experimentally realizable high-dimensional QKD.
\end{abstract}

\maketitle

\section{Introduction}
The fact that quantum mechanics allows for the secure distribution of cryptographic keys is one of the most significant insights within the field of quantum information \cite{bennettbrassard,Bennett92,BP93,barnett,nch,qkdreview}. An important subset of such quantum key distribution (QKD) protocols are those that use entanglement to securely distribute a key \cite{ekert,BBM92,NPWBK00}.  The security of entanglement-based QKD dependents on nonlocal correlations.  The basic idea is that an entangled state will be distributed between two parties, Alice and Bob.   The correlations between Alice and Bob's measurement outcomes can be used to characterize the shared state.  The actions of an eavesdropper inevitably alters the shared joint state.  By checking the correlations, Alice and Bob can determine if their shared state has changed.  The task of detecting an eavesdropper is thus a special case of characterizing a quantum state.  One should note, however, that a full state characterization is not necessary for QKD.  Instead, we are concerned with whether the states Alice and Bob act on are the intended ones.  There is thus a connection between entanglement-based QKD and experimental protocols that determine how well a  state can be generated \cite{stateestimate,statec1,statec2,statec3}.  

Increasingly, high dimensional entangled systems are being investigated for QKD.  These have two main advantages over more standard approaches, such as continuous variable QKD \cite{cvqkd}.  First, they allow multiple bits of secret key to be encoded on each photon.  In particular, it has been shown that, under realistic experimental conditions, it is, in principle, possible to encode over 10 bits per photon \cite{brougham}.  This means that high dimensional QKD protocols will be more efficient, in terms of bits per photon, than continuous variable protocols.  The second advantage is that high-dimensional systems can be more robust to certain types of noise \cite{CBKG,qudit1,qudit2}\footnote{There exist other sources of noise for which high-dimensional entangled states are less robust than low-dimensional states \cite{jon}.}.  

A simple and powerful way of encoding more information on each photon is to use time binning.  The essence of this approach is to use the photon's time of arrival as an extra degree of freedom onto which information can be encoded.  The possible arrival times can then be assigned to discrete time bins.  It has been demonstrated that photons can be prepared in states where their time of arrival is entangled \cite{tb1,tb2,tb3,tb4,tb5,BLPK}, which can be exploited for high-dimensional QKD \cite{etime,largealpha}.  The problem with such approaches is that, when the number of time bins becomes large, it can be difficult to make a measurement in a basis that is mutually unbiased with respect to the time of arrival.  This has led people to consider alternative ways to look for an eavesdropper, such as a Franson interferometer \cite{largealpha,franson}.  This device allows us to check the two time temporal correlation.  The basic idea is that an eavesdropper, referred to as Eve, will disturb this correlation and can thus be detected.

It is vitally important to note that Franson-based QKD protocols do not make measurements in two mutually unbiased bases.  This means that existing security proofs cannot be applied to them.  In this paper, we examine the security of Franson-based QKD schemes against a variety of eavesdropper attacks.  Our aim is {\it not} to obtain a full security proof for such protocols, but rather to analyse the security against specific attacks.  In particular, we consider attacks that localize the photons to multiple temporal locations.  We show that such attacks can severely compromise the security of protocols that use only one interferometer in each of Alice and Bob's setups.  Following this analysis, we explain how to modify the QKD protocols to mitigate against these attacks.  Our findings apply both to QKD protocols where the size of the time bins is fixed {\it a priori} and to protocols where this is not the case.

The outline of the paper is as follows.  In Sec. 2, we describe the basic idea behind time-bin-based QKD.  Section 3 presents a detailed examination of how Franson interferometers can be used in QKD protocols.  The action of an eavesdropper will be discussed in broad terms in Sec. 4.  This is followed in Secs. 5 and 6 by a more detailed treatment of the various eavesdropper attacks; in particular, Sec. 6 outlines the idea of attacks that have multiple temporal peaks.  Finally, the results are discussed in Sec. 7.

\section{High-dimensional time binned QKD}
Entanglement-based QKD is often implemented using photons that are entangled in their polarization \cite{scheidl}.  This limits each photon, however, to carrying at most one bit of secret key.  If we are to increase the amount of information encoded on each photon, we must use high-dimensional entanglement.  For example, one can use the orbital angular momentum of photons to encode multiple bits per photon \cite{jon,lesallen,gibson,kwiat}, though one is then susceptible to the deleterious effects of atmospheric turbulence in many applications \cite{boyd}.  
An alternative is to use photons that are entangled in their time of arrival, which is often called energy-time entanglement.  This type of entanglement can be  generated experimentally using the process of spontaneous parametric down conversion (SPDC) \cite{tb1,tb2,tb3,tb4,BLPK}. In order to describe energy-time entangled states, it is convenient to make use of continuous-time creation and annihilation operators, which satisfy the commutator relation $[\hat a(t),\hat a^{\dagger}(t')]=\delta(t-t')$ \cite{loudon}.  The two-photon entangled states can thus be written as
\begin{equation}
\label{gentimebin}
|\Psi\rangle_{12}=\int{dt_1dt_2g(t_1,t_2)\hat a^{\dagger}_1(t_1)\hat a^{\dagger}_2(t_2)|vac\rangle},
\end{equation}
where $\hat a^{\dagger}_{1,2}(t)$ is the creation operator either for photon 1 or 2, $|vac\rangle$ is the vacuum state and $g(t_1,t_2)$ is a normalized temporal envelope function that is zero when $|t_1-t_2|$ becomes sufficiently large.  Hence, if Alice, for example, detects a photon at time $t$, then Bob detects the other photon at a time close to $t$.  The photons are thus correlated in time, which can be exploited to distribute a shared string of randomly generated bits.  One way of doing this is to divide the photon's possible arrival times into discrete time bins and record whether photons are found in each time bin.  If we divide a single photon's arrival time into $M$ time bins, then the measurement record can then be used, in principle, to construct a string of $\log_2(M)$ bits.  

A practical time-bin encoding method that is amenable to efficient error-correction coding and can approach the quantum Shannon capacity of the channel uses contiguous time bins that form a `frame' \cite{framesqkd}.  Suppose we group the photon's possible arrival times into $M$ time bins, each with the same width $T$, forming the frame.  We label each of the $M$ time bins in the frame by an integer that runs from $0$ to $M-1$.  Detecting a photon in the $k$-th time bin means that the photon is localized to the time interval between $t=kT$ and $t<(k+1)T$.  Mathematically, this means that detecting the photon in the $k$-th time bin is equivalent to projecting onto the state 
\begin{equation}
|\tilde{k}\rangle=\int^{(k+1)T}_{kT}{\hat a^{\dagger}(t')|vac\rangle}dt'.
\end{equation}  
It is convenient to define the normalized states $|k\rangle=|\tilde{k}\rangle/\sqrt{T}$.  The states $\{|k\rangle\}$ form an orthonormal basis for the time-binned arrival time of a photon.  In practice, one might perform a timing measurement where the detectors have a resolution better than the spacing of each time bin.  One could then determine which time bin the photon was in and then discard the record of the exact time the photon was detected.  Such a measurement procedure is equivalent to projecting onto the basis states $|\tilde{k}\rangle$.   

The standard approach to ensuring the security of entanglement-based QKD is to introduce another measurement basis that is mutually unbiased with respect to the time of arrival. In particular, existing security proofs for $d$-level systems are all for the situation where one makes  measurements within mutually unbiased bases \cite{qudit1,qudit2}.  An example of basis states that are mutually unbiased with respect to the time of arrival, are the states
\begin{equation}
\label{mubs}
|\varphi_n\rangle=\frac{1}{\sqrt{M}}\sum_{k=0}^{M-1}{\exp\left(\frac{2\pi ink}{M}\right)|k\rangle},\;\;n=0,1,...,M-1,
\end{equation}
where $M$ is the number of time bins in each frame.  From an experimental perspective, measuring in the basis $\{|\varphi_n\rangle\}$ is challenging.  One possible approach is to use a linear optical network, as described in the Appendix.  This approach requires Alice and Bob to align $M-1$ Franson interferometers, which is very impractical when $M$ is large.  While it is difficult to measure in a basis that is mutually unbiased with respect to the time of arrival basis $|k\rangle$, it might be easier to make an approximate measurement. 

Due to the current difficulties in developing optical devices that can perform measurements in mutually unbiased bases, alternative approaches for securing time-bin-based QKD have been proposed.  One popular approach is to use a single Franson interferometer \cite{largealpha,shapiro}, whose operating principles are described in Sec. \ref{sec3}.  The security protocol is as follows.  Alice and Bob each have the optical setup shown in Figs. 1 and 2, where it can be seen that the source of the entangled photons is under Alice's control.  She will keep one photon out of each bi-photon pair for herself and send the other to Bob.  The photons that Alice keeps can be delayed by a controllable interval.  In the following analysis we assume that the time delay is equal to the time it takes for Bob's photons to arrive.  A photon incident on the variable beam splitter randomly selects between a path to detector $D_1$ or to pass through into the Franson interferometer.  Detection at $D_1$ corresponds to a measurement of the arrival time of the photon.  Detection at $D_2$ or $D_3$ correspond to checking for any possible disturbance caused by Eve (though obviously they could also yield some timing information, modulo the uncertainty of which way a given photon goes through the interferometer). 

The key idea behind the security check is that, if she tries to extract timing information, Eve will inevitable distrub the photon's temporal coherence between different times \cite{etime}.  The Franson interferometer allows Alice and Bob to look for disturbances in the temporal coherence and hence check for the presence of Eve.  It is important to note that Alice and Bob can only detect Eve from the instances when they both randomly choose to make a security check.  Similarly, Alice and Bob only obtain shared key bits when they both choose to measure the arrival time.  The instances when Alice and Bob make different measurements are discarded.  

\begin{figure}
\center{\includegraphics[width=8cm,height=!]
{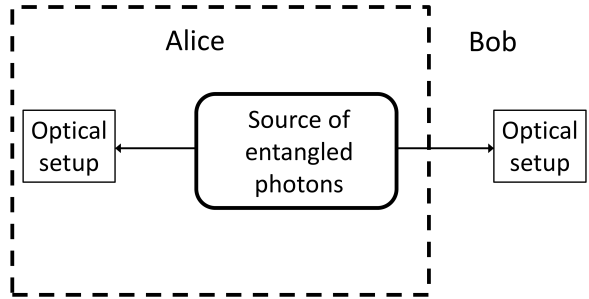}}
\caption{Schematic diagram of the high-level experimental arrangement for the QKD protocol.  An entangled photon source, which is controlled by Alice, produces entangled photon pairs.  One half of each photon pair is sent to Bob.  Alice and Bob then feed their photons into identical linear optical devices, as shown in figure 2.}  
\label{fig0}
\end{figure}

\begin{figure}
\center{\includegraphics[width=8cm,height=!]
{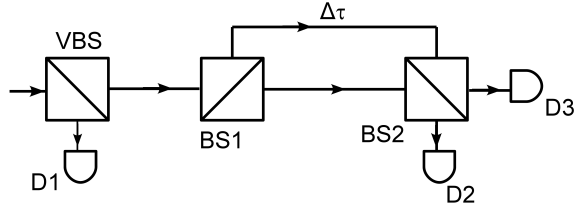}}
\caption{Schematic diagram of the optical apparatus used by Alice and Bob.  The element labeled VBS is a variable beam splitter, while BS1 and BS2 are both 50-50 beam splitters.  The elements $D_1$, $D_2$ and $D_3$ are all photon detectors.  The time difference between a photon in going along the long path instead of the short path, is $\Delta\tau$.}
\label{fig1}
\end{figure}

An extra layer of security can be achieved by using more than one Franson interferometer.  For example, two Franson interferometers can be used, each with different time differences between the interferometeric paths.  Alice and Bob then randomly choose which interferometer to use whenever they make a security check.  Conceptually, this is equivalent to using one interferometer and randomly switching between two different values for the time difference $\Delta\tau$.  Using multiple interferometers in this sense is quite different to what is discussed in the appendix, where the output of one interferometer is feed into the input of another.  To avoid confusion, we will only discuss the arrangement where we randomly switch between one of many interferometers.  The more complicated setup described in the appendix, will not be considered in the main text.  Another way of increasing the security is to check if Eve has substantially increased the errors in the photon time-of-arrival from what one would expect due to channel noise.  Alice and Bob can detect this by revealing a random sample of their time-of-arrival measurements and compare the observed error rate with the expected one.  

One key problem with multi-bit Franson-based QKD protocols is that a single Franson interferometer does not implement a mutually unbiased measurement.  This means that one cannot rely on existing security proofs to evaluate the security of a given protocol.  There do exist some heuristic arguments that suggest that an eavesdropper should always create a disturbance, which could in principle be detected by a Franson interferometer \cite{largealpha,shapiro}.  We will show, however, that their can exist attacks that cause very little disturbance, while extracting a significant number of timing bits.  Such attacks would be very difficult to detect.

\section{Detecting an eavesdropper using a Franson interferometer}
\label{sec3}
As the Franson interferometer is central to the QKD protocol, it is worth quickly summarizing how it acts on the incoming modes.  Let $\hat a(t)$ be the annihilation operator associated with the modes entering Alice's half of the interferometer, shown in Fig. 2.  In principle, Alice and Bob can use different values of the path differences.  We will, therefore,  differentiate between Alice and Bob's delays by adding a subscript $a$ or $b$.  The path difference in Alice's (Bob's) interferometer is thus $\Delta\tau_a$ ($\Delta\tau_b$).  As we are interested in time-binning the outputs, it makes sense to consider only values for $\Delta\tau_{a,b}$ that are integer multiples of the time-bin spacing.  It can be verified that the annihilation operators for the temporal modes incident on detector $D_2$ at time $t$ have the form 
\begin{equation}
\label{modes1}
\hat a_2(t)=\frac{1}{2}\left[\hat a(t)+\hat a(t-\Delta\tau_a)\right], 
\end{equation}
while the annihilation operators for the temporal modes incident on detector $D_3$ at time $t$ have the form 
\begin{equation}
\hat a_3(t)=\label{modes2}
\frac{1}{2}\left[\hat a(t)-\hat a(t-\Delta\tau_a)\right].
\end{equation}
If $\hat b(t)$ is the annihilation operator for Bob's modes, then the form of the modes incident at detectors $D_2$ and $D_3$, at time $t$, will have exactly the same form as Alice's, but with each $a$ changed to a $b$.  The form of these modes is crucial to the following security analysis.  The principal point is that Alice and Bob observe interference of the input modes from two different times at the Franson interferometer outputs.   

One can proceed with the analysis in full generality; the ideas are, however, best illustrated by looking at a specific example.  Suppose we have a source that naturally creates time-binned photons, such as a nonlinear crystal pumped by a mode-locked laser producing pairs of photons.  The entangled two-photon state can then be written in the form $|\Psi\rangle_{AB}=\sum_k{c_k|k\rangle_A|k\rangle_B}$.  As a further simplification, let us consider the case when the photons are equally likely to be found in any of the time bins.  The two-photon state is then given by 
\begin{equation}
\label{timebinstate}
|\Psi\rangle_{AB}=\frac{1}{\sqrt{M}}\sum_{k=0}^{M-1}{|k\rangle_A|k\rangle_B}.
\end{equation}
Alice and Bob can thus adjust the labeling of their time bins so that their $k$-th time bin corresponds to the time interval at which they would expect to detect photons emitted by, say, the $k$-th pulse. 

The form of the output modes shows that the effect of Alice obtaining a click at $D_2$, within her $m$-th time bin, is equivalent to projecting onto the state $|+,m,\Delta\tau\rangle$, while the effect of obtaining a click at $D_3$ within the $m$-th time bins is equivalent to projecting onto $|-,m,\Delta\tau\rangle$, where these states are defined as
\begin{equation}
|\pm,m,\Delta\tau\rangle=\frac{1}{\sqrt{2}}\left(|m\rangle\pm|m-\Delta\tau\rangle\right).
\end{equation}
It follows immediately that $\langle \pm,m,\Delta\tau|\langle \mp,m,\Delta\tau|\Psi\rangle=0$.  This means that, when Alice and Bob make a security check with $\Delta\tau_a=\Delta\tau_b$ and they both get clicks in the same time bin, then they must observe clicks at the same detector.  Hence, if Bob gets a click at $D_2$, Alice cannot observe a click at $D_3$ and vice versa.  The results for Alice and Bob's measurements are therefore correlated.  However, Eve can disturb the correlation.  Alice and Bob can thus check for an eavesdropper by looking at the correlation of their measurement results.  For example, suppose Eve makes a straight measurement of the arrival time of Bob's photon, i.e., {\it she projects onto the basis} $|k\rangle$.  The effect of this measurement is to completely destroy the correlation between the measurement outcomes for the Franson interferometer. 

In general, the two-photon state may not have the form of Eq. (\ref{timebinstate}).  Instead, we can be described the two-photon state using Eq. (\ref{gentimebin}).  The times $t_1$ and $t_2$ that appear in (\ref{gentimebin}) refer to the times the photons are emitted.  Suppose Alice detects a photon at time $t'$, which means that the photon was emitted at time $t'-\Delta T$, where $\Delta T$ is the time required for the photon to travel from the source to the detector.  To avoid having to talk of two different times, the emitted time and the detected time, we instead use only the emitted time.  This means that, if we say that Alice detects a photon at time $t_a$, then what we really mean is that the photon Alice detects was emitted at time $t_a$ and was detected by her at time $t_a+\Delta T$.  

When we construct the key, the only timing information that we require is the time bin which the photon is found within.  It might thus seem more natural to give our security analysis only in terms of time bins and not the continuous time variable corresponding to the emission time of the photon.  There are, however, some QKD protocols where the spacing of the time bins is not fixed a priori \cite{largealpha}.  In such schemes, one only decides on a time-bin spacing after Alice and Bob have established that there is no eavesdropper.  If our analysis is to include such protocols, then we must also discuss the security in terms of continuous time variables.

From Eqs. (\ref{modes1}) and (\ref{modes2}), we find that the probability for Alice and Bob to detect photons at the same detector is 
\begin{eqnarray}
\label{contdetect}
&&P^{AB}(D_j,D_j;t_a,t_b)=\nonumber\\
&&\frac{1}{4}\Big|g(t_a,t_b)+g(t_a-\Delta\tau_a,t_b-\Delta\tau_b)\nonumber\\
&&\pm g(t_a,t_b-\Delta\tau_b)\pm g(t_a-\Delta\tau_a,t_b)\Big|,
\end{eqnarray}
where the positive signs are for detection at $D_2$ and the negative signs are for detection at $D_3$.  The probabilities $P^{AB}(D_2,D_3;t_a,t_b)$ and $P^{AB}(D_3,D_2;t_a,t_b)$ are
\begin{eqnarray}
\label{contdetect2}
&&P^{AB}(D_i,D_j;t_a,t_b)=\\
&&\frac{1}{4}\Big|g(t_a,t_b)-g(t_a-\Delta\tau_a,t_b-\Delta\tau_b)\nonumber\\
&&+(-1)^jg(t_a,t_b-\Delta\tau_b)+(-1)^ig(t_a-\Delta\tau_a,t_b)\Big|,\nonumber
\end{eqnarray}
where $i\ne j$ and $i,j=2,3$.  Equations (\ref{contdetect}) and (\ref{contdetect2}) show that Alice and Bob's results will again be correlated.\footnote{In fact, the results could be correlated or anti-correlated, depending on the shape of $g(t_1,t_2)$.  For the sake of brevity we use the term correlated to refer to both cases.}  If an eavesdropper intercepts Bob's photon, she will cause the photons to be localized in time.  This means that the four detection probabilities will become approximately equal, {\it i.e.}, $P^{AB}(D_i,D_j;t_a,t_b)\approx P^{AB}(D_i,D_i;t_a,t_b)$.  The correlation in the measurement results will thus be destroyed.  In this way, Alice and Bob are able to detect an eavesdropper by checking the correlation in their measurements at $D_2$ and $D_3$.

\section{Modeling an eavesdropper attack}
As the photon source is under Alice's control, Eve can only intercept Bob's photons.  For the sake of simplicity we will not consider collective or coherent attacks \cite{qkdreview}; it is assumed that Eve only makes separate measurements on each photon.  Furthermore, we will also not consider attacks that add photons.  Nevertheless, if Eve is not careful, her attack will disturb the temporal correlation between Alice and Bob, which opens up the possibility of them detecting photons in different time bins.  From the perspective of extracting information, attacks such as this are suboptimal as they decrease the information shared between Alice and Bob.  Furthermore, it is easy for Alice and Bob to detect such attacks as all they need do is reveal a random sample of their timing information.  

A more interesting class of eavesdropper attacks are those that do not disturb Alice and Bob's temporal correlation.  These attacks are described mathematically using probability operator measures (POMs) \cite{barnett}, which are also called positive operator-valued measures (POVMs) \cite{peres}.  Consider first the special case in which the two-photon state is given by Eq. (\ref{timebinstate}).  If Eve is to extract information about Bob's photon without disturbing the temporal correlation, then her POM must have the form
\begin{equation}
\label{genpom}
\hat\Pi_k=\sum_{n}{\lambda_{k,n}|n\rangle\langle n|},
\end{equation}
where $\lambda_{k,n}\ge0$ for all of the values of $k$ and $n$.  The completeness condition for POMs, i.e., $\sum_k{\hat\Pi_k}=\hat 1$, implies that 
\begin{equation}
\sum_k{\lambda_{k,n}}=1,\;\forall\,n.
\end{equation}
A simple example is to set $\lambda_{k,n}=\delta_{k,n}$, which yields $\hat\Pi_k=|k\rangle\langle k|$, {\it i.e.}, a sharp time-of-arrival measurement.  

The previous formalism can be modified to apply to the general biphoton state given in Eq. (\ref{gentimebin}).  Consider first the case when Eve's attacks have a continuous set of possible outcomes.  The POM is then given by 
\begin{equation}
\label{contpom}
\hat\Pi(t)=\int{\beta(t;t')\hat b^{\dagger}(t')|vac\rangle\langle vac|\hat b(t')dt'},
\end{equation}
where $\beta(t;t')\ge 0$ for all values of $t$ and $t'$.  Furthermore, the completeness condition for POMs implies that $\int\beta(t;t')dt=1$.  One can also consider attacks that have a discrete set of outcomes.  This could correspond to Eve localize the photon to particular part of the frame.  The POM $\hat\Pi_{\mu}$ has a discrete index, which means that the envelope function changes to $\beta_{\mu}(t')$ and the normalization condition becomes $\sum_{\mu}{\beta_{\mu}(t')}=1$.

The simplest eavesdropper attack localizes Bob's photon to a single time bin.  In Sec. 3, we explain that such an attack disturbs the temporal coherence of the biphoton state.  This disturbance shows itself as a complete loss of correlation in the outputs of the Franson interferometer.  In order to avoid detection, Eve might use a measurement that does not localize the photon to only one time bin.  This will necessarily decrease the information that she is able to extract.  This  should, however, be balanced by the fact that the attack causes less disturbance to the temporal coherence of the two-photon state.

One can separate the possible attacks into two broad categories.  Attacks that localize the photons to several consecutive time bins and attacks that localize the photons to non-consecutive time bins.  The second type of attack can be thought of a filter function that has multiple temporal peaks.  

\section{Unsharp time of arrival attacks}
In this section, we will study attacks that localize photons to a band of consecutive time bins.  This approach can be thought as a time-of-arrival measurement where the resolution is not sufficient to determine the precise time bin of the photon.  

We first consider the case when the two-photon state has the form given in Eq. (\ref{timebinstate}).  Eve's attack is described by the POM
\begin{equation}
\label{unsharp}
\hat\Pi^L(k)=\sum_{r=kL}^{(k+1)L-1}{|r\rangle\langle r|},\;\;k=0,1,...,\frac{M}{L}-1,
\end{equation}
which localizes the photons to windows consisting of $L$ consecutive time bins.  The two-photon coherence is thus preserved over these $L$ time bins.  On average, this attack allows Eve to extract $\log(M)-\log(L)$ bits per intercepted photon.

It is clear that the correlation in Alice and Bob's results is completely destroyed by Eve's attack when $\Delta\tau_{a,b}>L$.  Consider instead the case when $L>\Delta\tau_a=\Delta\tau_b$ and where Eve localizes the photons over the time bins $kL$ to $(k+1)L-1$.  Furthermore, let Alice and Bob both have clicks in the $r$-th time bin.  When $r\ge kL-\Delta\tau_{a,b}$ and less than $(k+1)L+1$, Alice and Bob's results will be correlated and Eve will evade detection.  The eavesdropper can still be detected, however, when $L>\Delta_{a,b}$.  This follows from the fact that the POM (\ref{unsharp}) has sharp edges.  These sharp edges mean that Alice and Bob can detect Eve when they obtain clicks that are in time bins near the beginning or end of Eve's window.  A straightforward calculation shows that, for $L>\Delta\tau_{a,b}$, the probability for Alice and Bob to obtain different outcomes, when they both get a click in the same time bin, is
\begin{equation}
\label{windowdist}
P(A\ne B)=\frac{\Delta\tau_{a,b}}{2L}.
\end{equation}
As we explained, in the ideal setup with no eavesdropper, $P(A\ne B)=0$.  The probability $P(A\ne B)$ thus quantifies the errors introduced by Eve.  We will thus refer to $P(A\ne B)$ as being the probability for Eve to cause a disturbance, or equivalently, the error probability.  We see that the probability for Eve to disturb Alice and Bob's measurement correlation becomes small if $L$ is much larger than $\Delta\tau_{a,b}$.  However, as $L$ becomes larger, Eve extracts less information.  

The previous calculations reveal an important point.  To protect against an eavesdropper using an unsharp POM, such as (\ref{unsharp}), Alice and Bob should choose $\Delta\tau_{a,b}$ to be very large.  Alternatively, if two or more Franson interferometers are used, then at least one of the lengths $\Delta\tau_{a,b}$ should be large.  This will result either in Eve having a high probability of causing a disturbance, or she will be forced to make $L$ so large that the amount of information she gains is negligible.  In particular, to ensure that $P(A\ne B)$ is small, the amount of information that Eve extracts must be much less than $\log(M)-\log(\Delta\tau^*)$, where $\Delta\tau^*$ is the largest value for the path difference in Alice and Bob's interferometer.

\begin{figure}
\center{\includegraphics[width=8cm,height=!]
{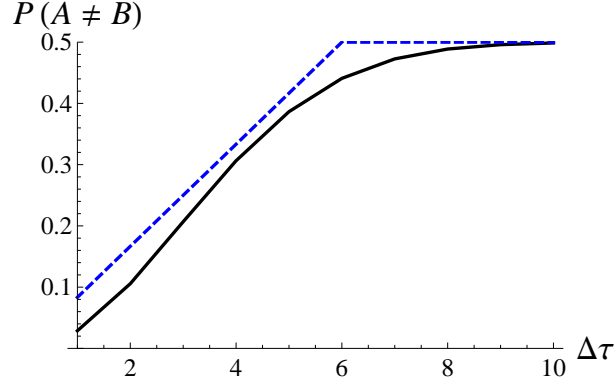}}
\caption{A plot of the probability for Alice and Bob to see different results, $P(A\ne B)$, against $\Delta\tau_{a,b}$.  The dotted line is the disturbance associated with a square window attack with $L=6$.  The solid line corresponds to an attack with a Gaussian window, where the width is such that it extracts the same information as the square attack with $L=6$. }
\label{fig2}
\end{figure}

The POM (\ref{unsharp}) is far from the only kind of attack capable of localizing photons to multiple consecutive time bins.  Alternative POMs can be found by using different choices for the coefficients $\lambda_{k,n}$ that appear in Eq. (\ref{genpom}).  For example, Eve can use a Gaussian, {\it i.e.}, $\lambda_{k,n}=\exp(-[k-n]^2/a)/Z_n$, where $Z_n=\sum{\exp(-[k-n]^2/a)}$.  This choice of window has the advantage of not having sharp edges.  This is balanced, however, against the fact a Gaussian is not flat over the range of time bins.  This can result in Alice and Bob observing some disturbance in their correlations for time bins near the middle of Eve's window.  In order to determine whether a Gaussian window is better than the square window used in (\ref{unsharp}), we compare the probability $P(A\ne B)$ for the case when both measurements extract the same information.  For example, we consider the case when $L=6$ and determine the width $a$, such that the Gaussian POM allows Eve to share the same mutual information with Bob that she would obtain by localizing the photons to 6 time bins.  Figure \ref{fig2} compares the disturbance caused by the Gaussian and square POMs.  It can be seen that the Gaussian POM causes less disturbance, not just for $L=6$, but for {\it all} other values.  Nevertheless, Fig. 3 shows that, when $\Delta\tau_{a,b}$ is sufficiently large, $P(A\ne B)$ becomes large enough for Eve to be readily detected.  We find that a Gaussian POM can be protected against by again making one of the values of $\Delta\tau_{a,b}$ large.  

The observations for the discrete case hold true also in the general case where the state is given by (\ref{gentimebin}).  Eve's POM is now described by Eq. (\ref{contpom}).  If Eve obtains the outcome $\hat\Pi(t)$, then the state is transformed to 
\begin{equation}
|\psi_t\rangle=\int\int{\beta(t;t')g(t_1,t')\hat a^{\dagger}(t_1)\hat a^{\dagger}(t')|vac\rangle dt_1dt'}, 
\end{equation}
which is a new two-photon state with the envelope function $\beta(t;t')g(t_1,t')$.  For the special case of $\beta(t;t')$ corresponding to a square window of width ${\mathcal L}$, one can verify that Alice and Bob's measurement correlation is completely destroyed when $\Delta\tau_{a,b}>{\mathcal L}$.  Furthermore, even when ${\mathcal L}>\Delta\tau_{a,b}$, the sharp edges of the window result in a non-zero probability for detecting Eve.  These problems cannot be remedied by using a different form for $\beta(t;t')$, {\it i.e.}, a general function with a single peak.  To see this, we note that $\beta(t;t')$ cannot be perfectly uniform if Eve's is to extract any information. Consequently, for a given value of $t$, the function $\beta(t;t')$ cannot have uniform support over the whole range of the frame and thus the envelope function for the two-photon state is changed.  In particular, the support of $\beta(t;t')g(t_1,t')$ will generally be over a smaller range than for $g(t_1,t')$.  Hence, the eavesdropper will disturb the coherence.  This can be detected by choosing one of the values of  $\Delta\tau_{a,b}$ sufficiently large.  The results of this analysis agree with earlier findings that show that Gaussian attacks decrease the Schmidt number of a two-photon state and thus can be detected by measuring the Schmidt number \cite{largealpha,kevin,shapiro}.  

\section{Eavesdropper attacks with multiple peaks}
In this section, we investigate eavesdropper attacks that have multiple temporal peaks.  The basic idea behind such attacks is that the Franson interferometer checks the temporal coherence between two times.  If Eve is to avoid detection, she must use an attack that preserves the coherence between several time bins.  Furthermore, the temporal spacing of these time bins should equal $\Delta\tau_{a,b}$.  The idea is best illustrated by looking at the special case where the biphoton state is given by Eq. (\ref{timebinstate}).  

Consider the multi-peaked attack that localizes the photons to $L$ time bins, given by
\begin{equation}
\label{mpgeneral}
\hat\Pi_k=\sum_{n=0}^{L-1}{\Gamma_{k,n}|k-n\Delta_e\rangle\langle k-n\Delta_e|}.
\end{equation}
To simplify the discussion, we set $\Gamma_{k,n}=1/L$.  If Eve obtains her $k$-th measurement outcome, then the two-photon state is transformed to 
\begin{eqnarray}
&&|\psi_k\rangle=\frac{1}{\sqrt{L}}\Big(|k\rangle_A|k\rangle_B+|k-\Delta_e\rangle_A|k-\Delta_e\rangle_B+...\nonumber\\
&&+|k-(L-1)\Delta_e\rangle_A|k-(L-1)\Delta_e\rangle_B\Big).
\end{eqnarray}
We assume that Eve sets $\Delta_e$ equal to $\Delta\tau_{a,b}$.  One sees that the temporal coherence is preserved for time bins that are not at the edges of $|\psi_k\rangle$.  For example, $\langle\pm,r,\Delta\tau|\langle\mp,r,\Delta\tau|\psi_k\rangle=0$, for $r=k$, $k-\Delta_e$,...,$k-(L-2)\Delta_e$.  For these time bins, Alice and Bob's results will be correlated and Eve will not be detected.  There will still be a disturbance if Alice and Bob detect photons from time bins at the edges of $|\psi_k\rangle$.  The important question is: how large is the probability of observing uncorrelated results?  Suppose Alice and Bob both observe a click in the same time bin; the probability for them to observe photons at different detectors is denoted as $P(A\ne B)$; this is the probability for Eve to introduce errors into the timing correlation.  A straightforward calculation shows that $P(A\ne B)=1/2L$ and that Eve extracts roughly $\log_2(M/L)$ bits of the key.  So, if $M=2^{10}$ and $L=32$, Eve can extract $5$ bits, while causing a disturbance of only $P(A\ne B)=0.0156$, which corresponds to a visibility of $97\%$ \footnote{We take the visibility to be the number of times Alice and Bob observe the same outcome minus the number of times they observe different outcomes, divided by the total number of outcomes, where all the outcomes are conditioned on Alice and Bob both having a detection within the same time bin.  The theoretical predicted visibility is just $V=1-2P(A\ne B)$.  The visibility is thus entirely equivalent to the probability of error.  It is used here to facilitate comparison with existing and future experimental results \cite{largealpha,darparep}.}.  Alice and Bob would thus need to achieve a Franson interferometer visibility greater than $97\%$ in their setup in order to protect against this attack.

\begin{figure}
\center{\includegraphics[width=8cm,height=!]
{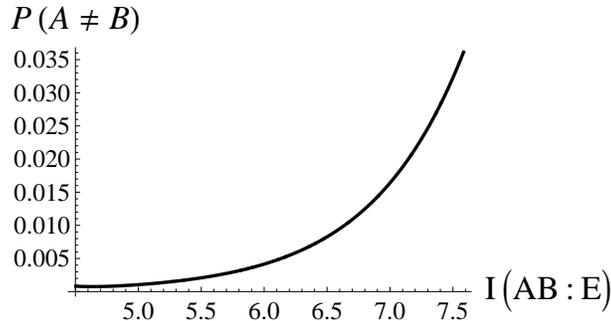}}
\caption{A plot of the probability that Eve introduces an error $P(A\ne B)$, versus the mutual information gain by Eve, $I(AB:E)$, for frames consisting of $M=2^{10}$ time bins.  Eve's attack is a multi-peaked Gaussian attack with $L=64$.}
\label{infocurve}
\end{figure}

By using a different choice for the coefficients $\Gamma_{k,n}$, Eve can construct a measurement that causes even less disturbance.  For example, she could use a Gaussian distribution $\Gamma_{k,n}=\exp(-\alpha[n-{L-1}/{2}])/\mathcal{N}_k$.  If $M=2^{10}$, $L=32$ and $\alpha=0.0335$, Eve can extract 6 bits while decreasing the visibility to $99.2\%$.  Similarly, for $L=64$ and $\alpha=0.0084$, Eve can extract 5 bits while decreasing the visibility to $99.8\%$.  These results can also be expressed in terms of the probability of error.  A typical information-disturbance curve for this particular attack is plotted in Fig. \ref{infocurve}, for $M=2^{10}$ and $L=64$.  The figure shows that Eve could extract about 6.5 bits while her probability of introducing an error is no grater than $0.01$.  While these attacks could in principle be detected, it would be very difficult from an experimental perspective.  The reason for this is that in a real experiment, the interferometer will not be perfectly aligned and the detectors will be inefficient and have dark counts.  These factors will result in $P(A\ne B)\ne 0$, even when there is no eavesdropper.  In previous experiments, the observed visibility was significantly below what is required to detect the outlined multi-peaked attack \cite{largealpha}. 

In all of these examples, we have assumed that Eve sets $\Delta_e=\Delta\tau_{a,b}$.  
There can, however, be cases where $\Delta\tau_{a,b}$ has multiple settings, {\it e.g}., Alice and Bob have two interferometers that they randomly switch between \cite{darparep}.  In this case, Eve has to guess the setting of $\Delta\tau_{a,b}$.  It can be shown that, if $\Delta_e\ne\Delta\tau_{a,b}$, then $P(A\ne B)=1/2$ for both of these attacks.  A disturbance of this size could easily be detected by Alice and Bob. 

The conclusion we draw from these results is that we should always use more than one Franson interferometer (or equivalently, randomly switch between more than one setting for $\Delta\tau$).  If only one interferometer with a single setting is used, then Eve can use a multi-peaked attack to extract a significant number of bits while causing very little disturbance.  However, if more than one setting for $\Delta\tau$ is used, then Eve must try and guess the value of used by Alice and Bob.  If she guesses incorrectly, then she destroys the correlation in Alice and Bob's results.  It is important to note that Alice and Bob can only check for an eavesdropper in the instances when they both use the same value for $\Delta\tau$.

The greater the number of possible values Alice and Bob have for $\Delta\tau$, the harder it is for Eve to guess the correct one.  The security of the QKD protocol is thus increased by having more values for $\Delta\tau$.  This can be made more precise by looking at how $P(A\ne B)$ scales with the number of settings for $\Delta\tau_{a,b}$.  Analytic results can be found for the simple attack where $\Gamma_{k,n}=1/L$.  If there are $d$ different values for $\Delta\tau$, the probability to obtain different outcomes for clicks within the same time bin is 
\begin{equation}
\label{proberrord}
P(A\ne B)=\frac{(d-1)L+1}{2dL}.
\end{equation}
Recall that $P(A\ne B)=0$ when there is no eavesdropper.  We see that the probability for Eve to introduce a disturbance increases as $d$ increases.  Furthermore, as $d$ tends to infinity, $P(A\ne B)$ tends to its maximum value of 1/2. 

It is instructive to take a different perspective on how using multiple settings for $\Delta\tau$ increases the security.  To detect Eve we would ideally want to measure in a basis $\{|\varphi_k\rangle\}$, which is mutually unbiased with respect to the time of arrival.  The more settings we have, the more information we gain about the temporal coherence between different times.  This will in turn give us more information about the possible measurement statistics for the basis $\{|\varphi_k\rangle\}$.  One could use the results for the Franson interferometers to estimate the hypothetical measurement statistics for the basis $\{|\varphi_k\rangle\}$.  The more settings we have for $\Delta\tau$, the better we would be able to estimate the statistics of the mutually unbiased measurement.  This implies that using multiple settings for $\Delta\tau$ will never increase the security beyond what one would have if we could measure in two mutually unbiased bases.  We thus find that increasing the number of settings for $\Delta\tau$ is {\it not} equivalent to increasing the number of measurement bases in a standard QKD protocol, such as that described in \cite{bruss,bourennane}.   

When Alice and Bob use multiple settings for $\Delta\tau$, Eve can adapt her attack in order to decrease the disturbance she induces.  One approach is for Eve to modify the POM (\ref{mpgeneral}) by changing the spacing between each peak.  Suppose Alice and Bob use $d$ incommensurate settings for $\Delta\tau$.  Eve can generalize the previous attack by localizing photons to a series of peaks where the spacing of subsequent peaks from the first one is of the form $n_1\Delta\tau_1+n_2\Delta\tau_2+...+n_d\Delta\tau_d$, with $n_1,n_2,...,n_d\in\{0,1,2,...,w-1\}$.  This attack localizes Bob's photon to one of $w^d$ time bins.  One POM that achieves this is 
\begin{eqnarray}
\label{wdpeak}
\hat\Pi^w_k&=&\frac{1}{w^d}\sum_{n_1=0}^{w-1}...\sum_{n_d=0}^{w-1}|k+n_1\Delta\tau_1+n_2\Delta\tau_2+...+n_d\Delta\tau_d\rangle\nonumber\\
&&\times\langle k+n_1\Delta\tau_1+n_2\Delta\tau_2+...+n_d\Delta\tau_d|.\nonumber\\
\end{eqnarray}
This attack allows Eve to extract roughly $\log(M)-d\log(w)$ bits per intercepted photon.  It can be shown that the probability for Alice and Bob to obtain a different outcome, when $\Delta\tau_a=\Delta\tau_b$ and they both obtain click in the same time bin, is
\begin{equation}
P(A\ne B)=\frac{1}{2w}.
\end{equation}
It is important to realize that this attack localizes the photon to $w^d$ time bins, not $w$.  In other words, Eve can only extract $\log(M)-d\log(w)$ bits.  While the attack (\ref{wdpeak}) is not optimal, it does illustrate the basic idea.  When multiple Franson interferometers are used, Eve can decrease her disturbance by localizing the photons to more time bins.  However, this means that she is able to extract less information.  By using a suitably large number of interferometers, we can ensure that she is able to extract only a negligible amount of the secret key.

\begin{figure}
\center{\includegraphics[width=8cm,height=!]
{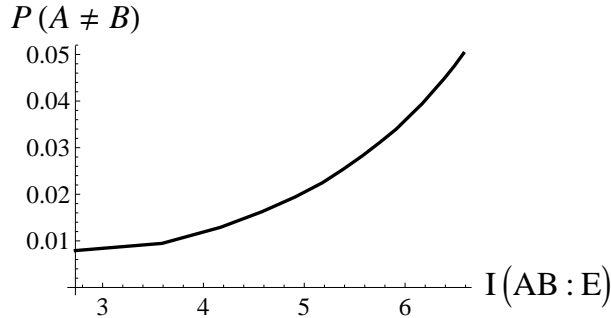}}
\caption{A plot of the probability of Eve introducing an error, $P(A\ne B)$ against the mutual information gain by Eve, $I(AB:E)$, for two Franson interferometers.  The plot is for frames consisting of $M=2^{10}$ time bins.  Eve's attack is a multi-peaked attack which is a product of two Gaussians, with $w=16$.}
\label{infocurve2}
\end{figure}

We can generalize the attack given in Eq. (\ref{wdpeak}) by introducing a weighting factor $\Gamma_{k,n_1,..,n_d}$ similar to in Eq. (\ref{mpgeneral}).  To illustrate how this can improve the attacks consider the example of Alice and Bob using two interferometers and trying to encode 10 bits per photon.  Eve can use a Gaussian weighting $\Gamma_{k,n_1,n_2}=\exp(-\alpha[n_1-{w-1}/{2}])\exp(-\alpha[n_2-{w-1}/{2}])/\mathcal{N}_k$.  For $w=16$ and $\alpha=0.3$,  Eve can extract approximately 5.1 bits while decreasing the visibility to $95.5\%$.  Alternatively, when $\alpha=0.2$, Eve can extract about 4.5 bits, while decreasing the visibility to $96.7\%$.  It is also useful to look at how the probability of error varies with Eve information gain.  This is plotted for $w=16$ in Fig. \ref{infocurve2}.  Comparing these results with those for a single Franson interferometer shows that using two Franson interferometers makes it harder for Eve to extract information without causing a noticeable disturbance.  Other attacks can be considered; a detailed study confirms that using more interferometers increases the security \cite{kevin}.  In particular, it can also be shown that, the amount of information Eve can extract decreases exponentially as the number of settings for $\Delta\tau$ increases \cite{kevin}.

We now explain how the previous results apply to the general two-photon state given in Eq. (\ref{gentimebin}).  An attack that localizes the photons to multiple temporal locations, such as (\ref{mpgeneral}), can be modeled using Eq. (\ref{contpom}), where the envelope function $\beta(t;t')$ has multiple peaks for the variable $t'$.  The simplest example is to use square windows of width $\delta$.  Eve's POM is then given by
\begin{equation}
\hat\Pi(t_e)=\sum_{m=0}^{L-1}{\int_{t_e-m\Delta_e}^{t_e-m\Delta_e+\delta}{\hat b^{\dagger}(t')|vac\rangle\langle vac|\hat b(t')dt'}}.
\end{equation} 
It may readily be verified that Alice and Bob's joint detection probability, $P^{AB}(D_i,D_j;t_a,t_b)$, are unchanged from Eq. (\ref{contdetect}) and (\ref{contdetect2}) when $\Delta_e=\Delta\tau_b$ and $t_e-(L-2)\Delta_e\le t_b\le t_e-\Delta_e+\delta$.  Detection within either of the intervals $t_e\le t_b\le t_e+\delta$ or $t_e-(L-1)\Delta_e\le t_b\le t_e-(L-1)\Delta_e+\delta$ will change the detection probabilities.  However, when $L$ is sufficiently large, the probability of detecting the photon within either of these edge time slots is small.  When $\Delta_e\ne\Delta\tau_b$, the detection probabilities are changed substantially and the correlation in Alice and Bob's measurement results is reduced.  We again see the need for more than one setting for $\Delta\tau$.  

The previous arguments can be extended to the situation where we have different shapes for the peaks.  For example, one can use Gaussians instead of square windows.  In this case, the calculation becomes more involved.  However, the same conclusion holds: if the spacing of the Gaussians equals $\Delta\tau_b$, then Eve can decrease her probability of detection.  This again demonstrates the need for more than one setting for $\Delta\tau$.  The other multi-peaked attacks, described by Eq. (\ref{wdpeak}), can also be studied in the general case.  All that is required is to choose a suitable form for the window function, $\beta(t';t)$, that appears Eq. (\ref{contpom}).  The specific results, such as the probability of error, will depend on the exact form of the two-photon state and the precise shape of $\beta(t';t)$.  The general conclusion, however,  that using more settings for $\Delta\tau$ makes it harder for Eve to avoid detection, still holds.  

\section{Conclusions}
We have investigated the use of Franson interferometers to secure high-dimensional time-bin-encoded QKD.  The approach was to encode information in the time of arrival of entangled photon pairs, one half of which would be sent to Bob.  If Bob's photons were intercepted by an eavesdropper, then the temporal coherence of the two-photon state would be disturbed.  Alice and Bob can detect this loss of coherence using a pair of Franson interferometers.  It is important to note that the setup does not utilize measurements within two mutually unbiased bases.  Hence, one cannot verify the security of Franson-based protocols by using existing methods.  Nevertheless, heuristic arguments have been presented that suggest that they should be secure \cite{largealpha,shapiro}.  The central aim of this work has been to investigate how secure this approach is with respect to {\it specific} attacks.  We do not present a general proof of security.  A key feature of our analysis was to consider eavesdropper attacks that have multiple peaks.  Such attacks allow Eve to extract information while preserving the coherence between several temporal points.  

One of the most important findings was to show that multi-peaked attacks allow an eavesdropper to extract significant information while causing little disturbance (see Fig. 4 for illustration of this point).  This has the immediate consequence that a single pair of Franson interferometers (one for Alice and one for Bob) is insufficient to secure the secret key against an eavesdropper.  Instead, one must use several different Franson interferometers, with different values for the length of the interferometeric path differences.  The error rate that is required to protect against these particular attacks was calculated, see Fig. 5.  We must stress, however, that the calculated values for the error probabilities (and visibilities) all assume a particular attack.  We have not presented a general security proof.  Nevertheless, these results are sufficient to show that a single pair of Franson interferometers is insecure and to strongly suggest that using multiple Franson interferometers can improve the security.

Several different types of multi-peaked attacks were considered.  For example, we looked at attacks that can be used against multiple Franson interferometers each with different settings for $\Delta\tau_{a,b}$.  It was demonstrated that for Eve to keep her disturbance small, the number of time bins she must localize the photons to scaled with the number of settings for the Franson interferometers .  This, in turn, greatly reduces the information that Eve can extract.

In all of the examples we have considered, Bob has a discrete set of possible choices for $\Delta\tau_b$.  An alternative situation is where Bob scans through different values for $\Delta\tau_b$.  This can be achieved if the Franson interferometer is setup in a Michelson-type arrangement \cite{largealpha}.  One might think that this enables Alice to only use one setting for $\Delta\tau_a$ while still ensuring security.  This is, however, not necessarily true.  The reason is that, if Bob's procedure for scanning through different values for $\Delta\tau_b$ is known, then Eve could use a multi-peaked attack where the value for $\Delta_e$ is modified in the same fashion as $\Delta\tau_b$.  Provided $\Delta_e$ is kept close to $\Delta\tau_b$, the disturbance caused will be minimal.  Furthermore, many entangled two-photon states have sharply peaked envelope functions $g(t_a,t_b)$.  This means that we are not sensitive to the disturbance caused by an eavesdropper when $\Delta\tau_b$ becomes sufficiently different from $\Delta\tau_a$.  Instead, the ability to detect a disturbance is confined to a small region of values for which $\Delta\tau_b$ is close to $\Delta\tau_a$.  Eve could thus simply use a multi-peaked attack with $\Delta_e=\Delta_a$.  We conclude that an essential requirement for using Franson interferometers in QKD is that {\it both} Alice and Bob have {\it multiple} different settings for the path differences of their interferometers. 

Collective measurements have not been considered in our analysis.  It is known that one can be secure against such attacks by performing a measurement in a basis that is mutually unbiased to the time of arrival \cite{qudit1,qudit2}.  Such a measurement can be made using an array of Franson interferometers as described in the Appendix.  Further work is needed, however, to determine if one could use fewer Franson interferometers and still detect such attacks.  What is clear is that a {\it single} pair of Franson interferometers (one for Alice and one for Bob) is vulnerable to certain attacks, although there are counter-measures that can be employed.  It is an open question whether the outlined modified setup suffices to uncover all possible attacks.

\section*{Acknowledgements}
This research was supported by the DARPA InPho program through the US Army Research Office award W911NF-10-0395.  SMB also acknowledges the Royal Society and the Wolfson Foundation for support.

\section*{Appendix: Using a linear network of Franson interferometers to implement a mutually unbiased measurement}
\begin{figure}
\center{\includegraphics[width=8cm,height=!]
{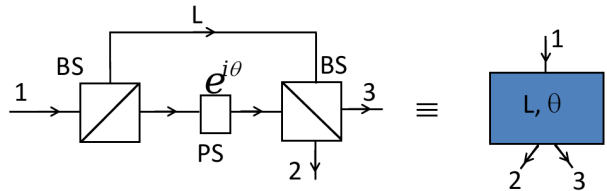}}
\caption{A schematic representation of the Franson interferometers on which the optical network is based.  BS1 and BS2 are both 50-50 beam-splitters, while PS represents a phase shifter.  The numbers $\theta$ and $L$ are the phase of the phase shifter and the path difference between the long and short paths of the interferometer, respectively.}
\label{figap1}
\end{figure}

\begin{figure}
\center{\includegraphics[width=9cm,height=!]
{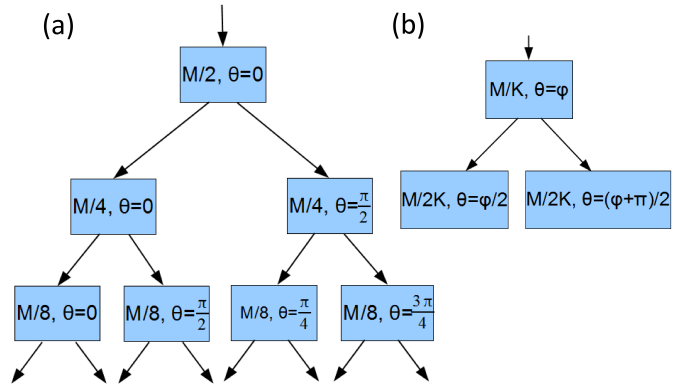}}
\caption{A schematic diagram of the optical network that Alice and Bob each uses to measure in the basis $\{|\varphi_k\rangle\}$.  Each block is a Franson interferometer, which is shown in Fig. \ref{figap1}.  Figure (a) shows the network when we have 8 time bins.  Detection of a photon at one of the outputs corresponds to projecting onto one of the basis states $|\varphi_k\rangle$.  Figure (b) shows the general rule for constructing each new row of the network.  A network can be designed that secures $2^N$ time bins, by repeated use of this rule.}
\label{figap2}
\end{figure}

In this Appendix, we outline how one might measure in the basis defined in Eq. (\ref{mubs}).  The approach uses only linear optics and can be thought of as a multiport version of a Franson interferometer.  The path difference between the long and short paths of each interferometer is an integer multiple $L$ of the time bin widths.  It is convenient to represent each of the Franson interferometers schematically as blocks with one input and two outputs, as shown in Fig. \ref{figap1}.  Each block can be characterized by two numbers, $L$ and $\theta$, where $\theta$ is the phase of the phase shifter.  Alice and Bob each have the optical network shown in Fig. \ref{figap2}.  Detection of photons at one of the outputs corresponds to projecting onto one of the basis states $|\varphi_k\rangle$.  

If we divide each photon's time of arrival into $2^N$ time bins, we require $2^N$ outcomes for any measurement that is mutually unbiased to the time of arrival.  This means that Alice and Bob's networks each have $2^N$ outputs and thus each require $2^N-1$ interferometers.  The total number of interferometers thus scales linearly with the number of time bins (or, equivalently, the dimensions of Alice and Bob's Hilbert space) \cite{zeilinger}.  However, the number of bits that can be encoded on each photon is not $2^N$ but $N$.  The number of interferometers thus scales exponentially with the number of bits per photon.  Furthermore, stabilizing the setup, even for low number of bits per photon, would be challenging.  We conclude that the proposed setup is not a practical means of securing high-dimensional time-bin-based QKD.

An alternative approach is to use only one branch of the full tree shown in Fig. \ref{figap2}.  In this case each output of a branch that feeds into a different branch of the tree will, instead, be terminated by going to a detector.  This allows Alice and Bob to each project onto one of the states $\{|\varphi_k\rangle\}$.  We expect that this approach is not as secure as using the whole tree.  Nevertheless, it might still be useful in detecting certain attacks.  A full analysis of this setup is beyond the scope of the current article.

\end{document}